\documentclass[12pt,a4paper]{article}
\usepackage{graphicx}

\begin{document}

\title{\bfseries Classroom computer animations of relativistic objects}
\author{Leo Brewin\\
School of Mathematical Sciences\\
Monash University\\
Australia}

\date{26-July-2003}

\maketitle

\begin{abstract}
This is a short note to announce the availability of some movies that may be
useful in classroom discussions on the photographic appearance of objects
moving at relativistic speeds. The images are based on special relativity with
no account taken of (other than to ignore) the effects of doppler shifts,
intensity shifts or gravitational effects. 
\end{abstract}

{\bfseries\large Motivation}

First encounters with Special Relativity can be quite daunting for many
students: time dilation, length contraction, twin paradoxes and the like all
tend to send their minds into a state of neural meltdown. This is not a
favorable outcome. But they do respond well to colour and movement -- in
particular to computer simulations of special relativistic effects such as the
apparent change in shape, as revealed by a photograph, of an object moving
close to the speed of light. I have made a series of movies for a number of
common situations using a (slightly) modified version of the standard
ray-tracing package, POV-Ray(tm) 3.5. Though there is nothing new in this, I
have found it to be a useful teaching tool. I have put the movies and other
material on the web at this address

{\tt http://www.maths.monash.edu.au/$\sim$leo/relativity/sr-photography}

The first point that must be stressed before taking the students through
these movies is that the process of taking a photograph in Special Relativity
is fundamentally different from what we normally call making an observation.
Photography entails the collection of photons by one person at one instant in
time (in the camera's frame) while an observation is the detection of a
single event. Furthermore, an observer (or a frame) is the infinite arena in
which all possible events can be detected.

I found the material in these movies sufficient to occupy approximately two
one hour tutorials. The students can be asked to study the movies and
then to explain, both in mathematical and descriptive terms, why the images
are as they are. In doing so they will have to have a firm understanding of
the principle of relativity and of relativistic aberration. My aim was to
get the students to formulate a simple principle that would help them
predict how an object's shape might appear when photographed at relativistic
speeds.

I have not included the effects of redshift nor the changes in intensity.
I was interested only in the apparent change in the shape rather than changes
in colour and brightness (perhaps at later time, any volunteers?).

The changes to the POV-Ray(tm) package were quite simple. I added a few routines
to allow Lorentz transformations between the camera and object frames. The
camera was allowed to move relative to all other objects in the scene. All of
the photons were created in the rest frame of the camera. These were then
Lorentz transformed into the scene frame at which point the normal POV-Ray(tm)
routines takeover. This idea has been used before (see for example the code
by Andrew Howard {\tt andrbh@cs.mu.oz.au} though I believe there are errors in
his Lorentz routines).

I have included the movies and the modified POV-Ray(tm) source at the above
website. Note, as part of the POV-Ray(tm) licence I am required to declare that
my modifications do not form part of the official POV-Ray(tm) release and that
I take full responsibility for all alterations. Furthermore, if you wish to use
this version you must read and accept the legal conditions as set out in
povlegal.doc (the latest version of which can be found at
{\tt http://www.povray.org/povlegal.html})

\vskip 0.25\parskip

{\bfseries\large Forward motion - The Triffid Nebula}

Here are two (computer generated) images (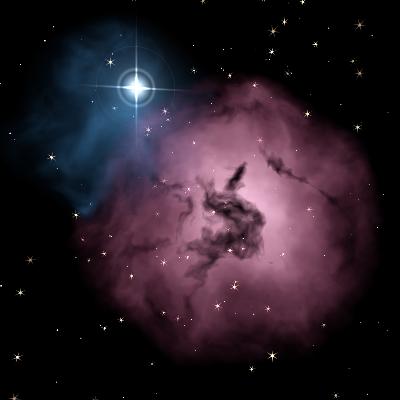, 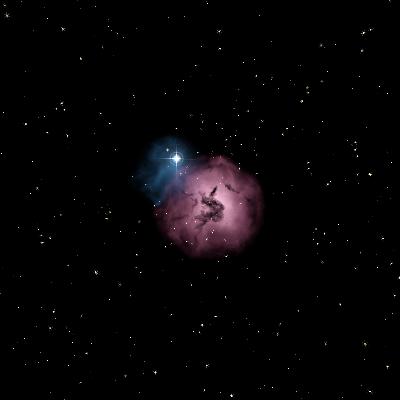) of the Triffid
Nebula. Both were taken with identical cameras.  The only difference is that
one of the images was taken in a frame moving at $\beta=0.75$ towards the
Nebula. The images were taken at the instant the two cameras passed each
other. Which image is which? Compare this with the familiar images from
Startrek when the warp drive is engaged.

\begin{center}
\framebox[0.6\textwidth]{{Figure 1a}\vrule height 0.3\textwidth depth 0.3\textwidth width 0pt}

\framebox[0.6\textwidth]{{Figure 1b}\vrule height 0.3\textwidth depth 0.3\textwidth width 0pt}
\end{center}

\clearpage

{\bfseries\large Forward motion - The TARDIS}

Here are another two stills (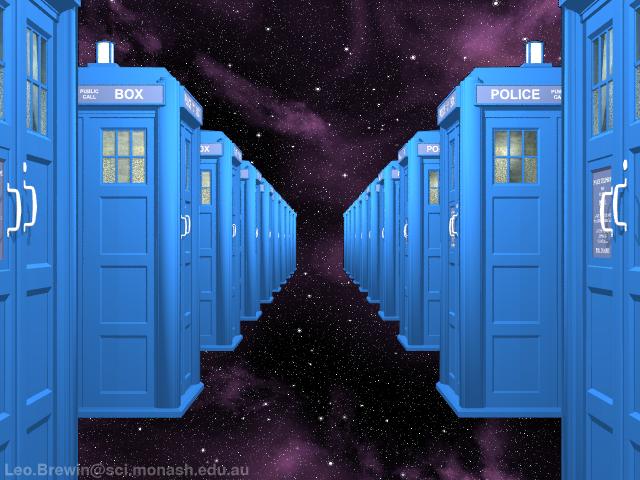, 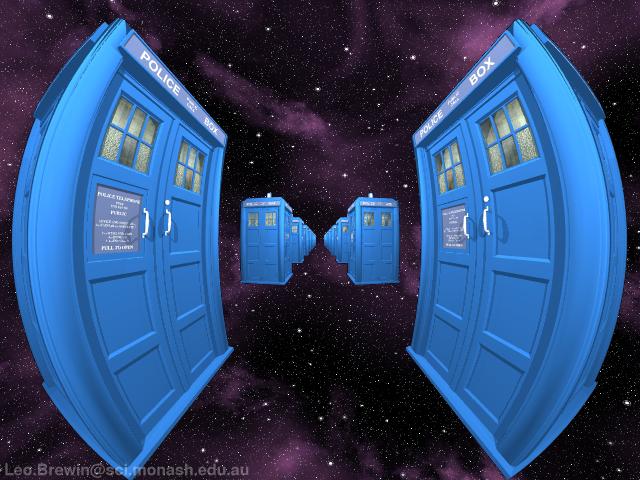), this time of a cube (Dr
Who's \emph{TARDIS = Time and Relative Dimensions in Space}) moving directly
towards the observer. The cubes appear to bulge outwards. Why? They also
appear to be smaller (or are they further away?). Why? 

\begin{center}
\framebox[0.8\textwidth]{{Figure 2a}\vrule height 0.3\textwidth depth 0.3\textwidth width 0pt}

\framebox[0.8\textwidth]{{Figure 2b}\vrule height 0.3\textwidth depth 0.3\textwidth width 0pt}
\end{center}

\clearpage

{\bfseries\large Transverse motion - The Albert Einstein Steam Railway}

Here are two images of a train moving across the line of sight of a single
observer. The top image (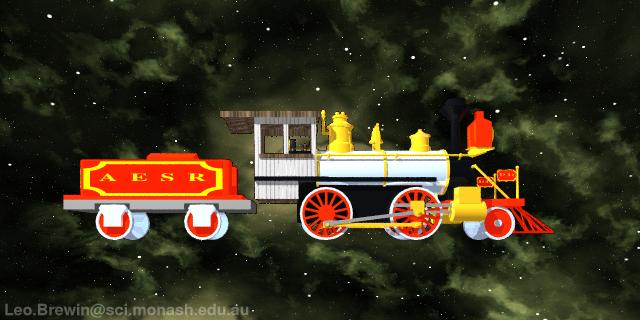) is of a train moving at $\beta=0.05$
while the lower (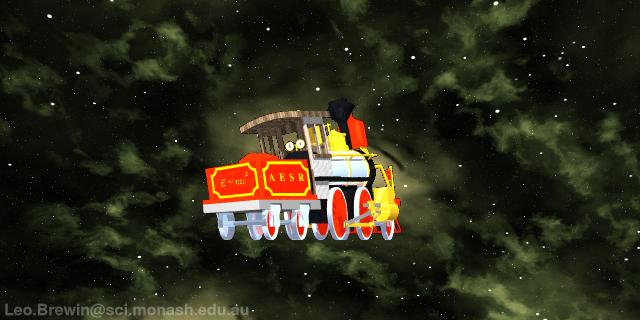) image has $\beta=0.95$.

The students should be able to explain why speed of the train appears to slow
down as it moves from left to right. The train also appears to shorten in
length as it moves left to right. Why? The train also appears to rotate, but
that is harder to explain (bonus marks?)

\begin{center}
\framebox[0.95\textwidth]{{Figure 3a}\vrule height 0.25\textwidth depth 0.25\textwidth width 0pt}

\framebox[0.95\textwidth]{{Figure 3b}\vrule height 0.25\textwidth depth 0.25\textwidth width 0pt}
\end{center}

\end{document}